\documentclass{article}

\usepackage{arxiv}
\usepackage{graphicx, color}
\usepackage{listings}
\usepackage{multirow}
\usepackage{amsmath}
\usepackage{ifthen}
\usepackage{program}
\usepackage{subscript}
\usepackage{spverbatim}
\usepackage[linesnumbered,ruled,vlined]{algorithm2e}

\newcommand{\HIDE}[1]{} 
\newcommand\david[1]{\nb{David}{#1}}

\let\oldnl\nl% Store \nl in \oldnl
\newcommand{\nonl}{\renewcommand{\nl}{\let\nl\oldnl}}% Remove line number for one line


\newtheorem{definition}{Definition}

\usepackage{listings}
\usepackage{color}

\definecolor{dkgreen}{rgb}{0,0.6,0}
\definecolor{gray}{rgb}{0.5,0.5,0.5}
\definecolor{mauve}{rgb}{0.58,0,0.82}

\title{An Empirical Study on Failed Error Propagation in Java Programs with Real Faults}

\author{
  Gunel Jahangirova\\
  Università della Svizzera italiana\\
  Lugano, Switzerland \\
   \And
 David Clark \\
  University College London\\
  London, UK\\
  \And
 Mark Harman \\
  Facebook \& University College London\\
  London, UK\\
  \And
  Paolo Tonella\\
  Università della Svizzera italiana\\
  Lugano, Switzerland \\

}

\begin{document}
\maketitle

\begin{abstract}
During testing, developers can place oracles externally or internally with respect to a method. Given a faulty execution state, i.e., one that differs from the expected one, an oracle might be unable to expose the fault if it is placed at a program point with no access to the incorrect program state or where the program state is no longer corrupted. In such a case, the oracle is subject to failed error propagation. 

We conducted an empirical study to measure failed error propagation on Defects4J, the reference benchmark for Java programs with real faults, considering all 6 projects available (386 real bugs and 459 fixed methods). Our results indicate that the prevalence of failed error propagation is negligible when testing is performed at the unit level. However, when system-level inputs are provided, the prevalence of failed error propagation increases substantially. This indicates that it is enough for method postconditions to predicate only on the externally observable state/data and that intermediate steps should be checked when testing at system level. 

\end{abstract}

\section{Introduction} 

%%%%%%%%%%. Mark's additions %%%%%%%%%%%%%%%%%%

Software faults are difficult to detect and fault removal consumes a significant proportion of software development and evolution~\cite{beizer:testing,bertolino:testing-fose07}.
One of the widely-attributed sources of such difficulty is the possibility of \textit{failed error propagation} (FEP): 
%% Note: grottke:classification (http://www.grottke.de/documents/AClassificationOfSWFaults.pdf (this paper explicitly talks about complexity of failed error propagation) 
%% cites Adaptive Control Approach for Software Quality Improvement
%% On Failures and Faults, Brian Randell, 
%% Service Availability: Principles and Practice - book 
a fault may corrupt the program's internal state, yet this corruption fails to propagate to any point at which it is observed \cite{grottke:classification,randell:fandf,wong:adaptive}. Such non-propagating faults play the role of `nasty unexploded mines'; lurking undetected in software systems, waiting for that slight change in execution environment that allows the corrupted error state to propagate, causing unexpected system failure.

Despite the importance of FEP, surprisingly few empirical studies in the literature assess the extent of the problem. Empirical evidence based on a few examples of real faults is available only for the C/C++ programming language~\cite{setTimeLimit,chandra:how,gunawi:eio,DaranT96}, while for Java results have been obtained only with mutants~\cite{MasriA14,XiongH0ZZL15}, not with real faults. In the absence of robust empirical analysis, the research and practitioner community is left with suspicions of a silent menace of unknown proportions.

In order to bridge the gap between suspicions and empirical evidence we set out to perform a large empirical study of FEP on real faults from Defects4J~\cite{just:defects4j}, a large scale benchmark that has become the de-facto standard~\cite{Just:2014:MVS,Ma:2015:GRT,B.Le:2016:LBF,Lu:2016:RTP,Mart:2016:APR,Xin:2017:ITP,Pea:2017:EIF,Xiong:2017:PCS,Sham:2015:AUT} for real faults in Java programs. Our study encompasses all six projects in Defects4J and the associated 386 real faults. 
%The goal of this paper is to investigate empirically the occurrence of FEP when real faults are taken into account. 
Since FEP is a statistical property of a method (in fact, it may occur in some executions and not in others), we faced the problem of obtaining a sample of empirical data that is large enough to draw statistically meaningful conclusions. This requires that the considered faults are executed multiple times, in program executions that differ from each other, and that the effects of the faults on the program state are observed along corresponding execution points of both faulty and fixed program.  We have extended the EvoSuite~\cite{fraser:evosuite}  test case generator to address the first problem and we have developed our own trace alignment algorithm to address the second problem.
%Hence, we also had to align the two executions, in order to be able to compare the execution states at comparable execution points.
%The benchmark used in the empirical study is Defects4J~\cite{just:defects4j}. In particular, we considered 6 Java projects, 395 real bugs from these projects and 536 methods changed to fix such bugs.

Our study revealed a very surprising finding: in this significant corpus of real world bugs, the prevalence of unit-level FEP is negligible. We further experimented with seeded synthetic faults (mutants \cite{yjmh:analysis}), for which we observed that unit-level FEP {\em was} found to be much more prevalent. To further analyse the propagation of real faults we conducted experiments testing programs at the system level rather than on unit level. Our results show that the rate of system-level FEP with real faults is substantially higher than the rate of unit-level FEP both with real and synthetic faults.

\section{Failed Error Propagation}\label{sec:fep}

The effectiveness of testing depends on the use of oracles that are sensitive to any deviation from the intended program behavior and that report all such deviations as test failures. One of the key decisions about the use of oracles is their placement. Oracles can be placed in test cases in the form of a test case assertion, i.e., outside the method under test (unit level testing) or at the end of the entire system execution (system level testing); at the end of the execution of the method under test, before the return point (acting as a post-condition); or even internally, at any arbitrary execution point, predicating on the intermediate program states observed during method execution. 

An external/output oracle (i.e., a test case oracle) has limited capability to discriminate between incorrect and correct method executions, since it can only check the value returned by the method under test and the externally observable state affected by the method under test (e.g., global variables, externally observable object states, persistent chan\-ges in the environment, output produced by the whole system execution). In a specific program execution, an error may escape detection by an external/output oracle if it generates an internal state that differs from the expected one without producing any externally visible effect. This means it returns the expected value and it changes the externally visible state in the expected way. Of course, in order for this to be an error, there must be at least one execution where the error produces an externally visible incorrect effect. Hence, external/output oracles can eventually detect all faults, but they might require a lot of test cases if there is only a low probability that the internal state differences propagate to externally visible differences. When this happens at the unit level, we say the method is subject to \textit{external failed error propagation} (\textit{extFEP}). When this happens at the system level, we say the method is subject to \textit{system failed error propagation} (\textit{sysFEP}). External/output oracles are weak in comparison with return point or internal oracles when external/system FEP happens.

At the unit level a return point oracle (i.e., an internal oracle placed right before the return point) is more powerful than an external oracle because it can predicate on the entire execution state at the return point, not just on the externally visible state. However, return point oracles may also be subject to FEP -- in this case, called \textit{internal FEP} (\textit{intFEP}). In fact, in a specific program execution, the error, which we assume as detectable externally in other executions, might generate an internal state which differs from the expected one, but such a difference might disappear when the execution proceeds from the faulty statement to the return statement, where no state difference with respect to the expected state is observed. 

\begin{figure}[htb]
\begin{lstlisting}
int f(int x) {
    // pp0: assert(\old(x) == x))
    x = 3 * x; // fix: x = 2 + x;
    // pp1: assert(\old(x) + 2 == x))
    if (x > 0) {
        // pp2: assert(2+\old(x) > 0 && \old(x)+2 == x))
        x = x % 4; 
        // pp3: assert(2+\old(x) > 0 && (\old(x)+2) % 4 == x))
    } else {
        // pp4: assert(2+\old(x) <= 0 && \old(x)+2 == x))
        x = x + 1; 
        // pp5: assert(2+\old(x) <= 0 && \old(x)+3 == x))
    }
    // pp6: assert(2+\old(x) > 0 ? \result == (2+\old(x)) % 4 : \result == 3+\old(x));
    return x;
}
void test0() {  assert(f(4) == 2); } // FAIL
void test1() {  assert(f(5) == 3); } // PASS
\end{lstlisting}
\caption{Code example including 7 possible internal oracle placement points, \texttt{pp0} to \texttt{pp6}, as well as a test case (\texttt{test0}) exhibiting no FEP and one with external FEP (\texttt{test1})}
\label{fig:example}
\end{figure}

%A program is subject to \textit{failed error propagation} if a fault at a given program point changes the program execution state at such point, with respect to the execution of the fault-free program, but the faulty execution state does not propagate to any existing (possibly implicit) assertion that reveals the error (in mutation analysis, this is the case of a mutant that is not killed). Hence, the error fails to propagate from the faulty statement to a program point where an assertion can reveal it (the implicit assertions, which crash the program upon error, are included among the existing assertions). 

In the running example shown in Figure~\ref{fig:example}, consider the faulty statement \verb"x = 3 * x", whose corresponding fixed version is \verb"x = 2 + x", and the return point assertion at \texttt{pp6}. If the faulty program is executed with input \texttt{x==4} (see \verb"test0" in Figure~\ref{fig:example}), it returns 0, while the expected value is 2, which indicates the fault can indeed affect an externally visible result, in some execution. If the program is executed with input \texttt{x==5}, we can observe a different execution state at program points \texttt{pp1} and \texttt{pp2}, where we have \texttt{x==15} in the faulty program, while we expect \texttt{x==7}. However, at program point \texttt{pp3} the same value of \texttt{x} is produced by both the faulty and the fixed program: \texttt{x==3}. When the assertion at \texttt{pp6} is executed, no difference is observed between faulty and fixed program. The external assertion inside \verb"test1" also does not fail. 

In the second program execution (with input \texttt{x==5}), the error fails to propagate to the assertion at \texttt{pp6} because the information about the different execution states in the faulty and fixed programs is destroyed by the execution of statement \texttt{x = x \% 4}, which collapses the two different program states into the same one, \texttt{x==3}. This is a case of both internal and external FEP, which could be solved by introducing the internal assertion at \texttt{pp1} or \texttt{pp2}.

Consider a case where external/output FEP occurs, while internal FEP does not.
Suppose we change the return type of \texttt{f} in the example shown in Figure~\ref{fig:example} to \verb"boolean" and change the return expression to \verb"(x >= 0)". With such a change, \verb"test0" would pass, expecting and observing \verb"true" as return value. However, the return point assertion at \texttt{pp6} would fail, since the observed value \texttt{x=0} differs from the expected value \texttt{x=2}.

\begin{definition}[Internal/External/System FEP]
Given a fault $f$ at program point $pp_f$, a specific method execution $e$ containing $pp_f$, represented as the sequence of program points $e = \langle pp_0, \ldots, pp_n\rangle$, is said to be subject to internal/external/system failed error propagation (intFEP, extFEP, sysFEP) if execution of the faulty statement $pp_f$ causes a state divergence between actual and expected execution states, $s[pp_f]$ and $s'[pp'_f]$, which is not observable respectively at the return statement $pp_n$ (intFEP), outside the faulty method (extFEP) or in the output produced by the system (sysFEP):
\begin{align*}
\textstyle{intFEP}: & s[pp_f] \neq s'[pp'_f] \wedge s[pp_n] = s'[pp'_n] \\
\textstyle{extFEP}: & s[pp_f] \neq s'[pp'_f] \wedge ext = ext' \\
\textstyle{sysFEP}: & s[pp_f] \neq s'[pp'_f] \wedge out = out'
\end{align*}
where program points $pp'_f, pp'_n$ correspond to $pp_f, pp_n$ in the fixed program; $s$ and $s'$ indicate the execution state of faulty and fixed program respectively; $ext$ and $out$ are the values output by the unit and the system respectively.
\end{definition}

\begin{figure}[htb]
\centering
\includegraphics[width=9cm]{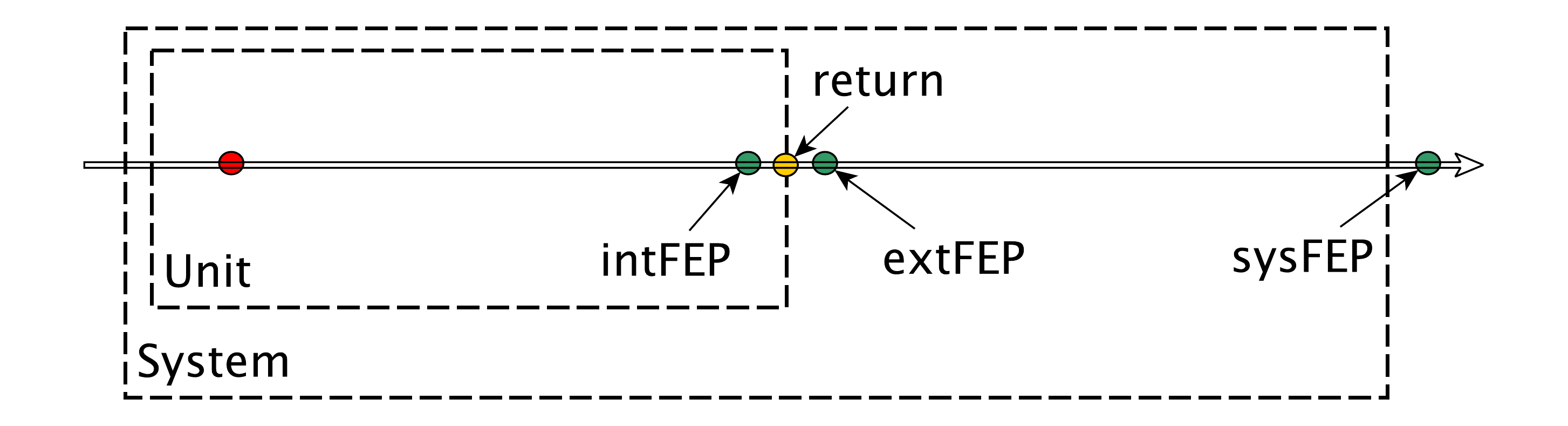}
\caption{Execution points where state corruption disappears in case of internal, external or system FEP}
\label{fig:fep-plot}
\end{figure}

\begin{table*}[htb]
\caption{Studies on Failed Error Propagation}
\label{tab:fep_studies}
\begin{small}
\begin{center}
\begin{tabular}{|l|r|r|r|r|r|r|r|}
\hline
\textbf{Study} & \textbf{Language} & \textbf{Fault Type}  & \textbf{Number of Faults} & \textbf{FEP Type} & \textbf{FEP Ratio}\\
\hline
Daran et al.~\cite{DaranT96} &  C &  Real & 12 & FEP & 78\%\\
\hline
Masri et al.~\cite{masri2009empirical}, Masri \& Assi~\cite{masri2010,MasriA14} & Java, C &  Seeded & 148 &  CC, FEP & 56.4\%, 15.7\%\\
\hline
Wang et al.~\cite{WangCCZ09} &  C & Mutants  & 3000 & CC & 36\%\\
 &   & Real & 38 & CC  & 53.75\%\\
\hline
Miao et al.~\cite{miao2012} & C & Seeded & 115 & CC & 56.26\%\\
\hline
Li \& Liu~\cite{li2012using} & C & Seeded & 18 & CC & 20.4\% \\
\hline
Xue et al.~\cite{XuePN14} & Java & Seeded & 19 & CC & 7.4\% \\
\hline
Androutsopoulos et al.~\cite{AndroutsopoulosCDHH14} & C & Mutants & 1408 & CC, FEP & 4.89\%, 9.85\% \\
\hline
Xiong et al.~\cite{XiongH0ZZL15} & Java & Mutants & 137 & FEP & 43.4\% \\
\hline
This paper & Java & Real Faults & 386 & FEP (Unit-level) & 0\%\\
 &  & Mutants & 1125 & FEP (Unit-level) & 3.7\%\\
  &  & Real Faults & 132 & FEP (System-level) & 11.4\%\\
\hline
\end{tabular}
\end{center}
\end{small}
\end{table*}

It can be easily shown that internal FEP subsumes external FEP, which in turn subsumes system FEP (\textit{intFEP} $\Rightarrow$ \textit{extFEP} $\Rightarrow$ \textit{sysFEP}). Figure~\ref{fig:fep-plot} shows the three cases of FEP in graphical form. When a state corruption occurs in the execution (red leftmost dot), it might no longer be observable in the execution right before the return point (green dot labeled intFEP), right after the return point (extFEP dot) or when the entire system execution is over (sysFEP dot). Visually, the subsumption relation corresponds to the green dot (second from left) in the figure, which propagates from left to right (i.e., if a state corruption disappears, it remains unobservable until the end of the execution).

\section{Background and Related Work}\label{sec:background}

\subsection{Sensitivity Analysis}

Voas and Jeffrey~\cite{Voas1992} introduced the Propagation, Infection, Execution (PIE) framework to estimate three probabilities: 1) the probability that a particular section of a program is executed, 2) the probability that the executed section affects the data state, and 3) the probability that the affected data state has an effect on the program output. This approach was also reiterated in the Reachability-Infection-Propagation (RIP) model described in Amman and Offutt~\cite{AmmannOffutt2008}. 

Based on the PIE framework, Voas et al.~\cite{Voas1991} proposed a testability technique called "sensitivity analysis", which is the process of determining the sensitivity of a location in a program. Here the word "sensitivity" means a prediction of the probability that a fault will cause a failure in the software at a particular location under a specified input distribution. Sensitivity analysis estimates the probability that a code location will be executed, the probability that faults will cause data-state errors, and the probability that data-state errors will change the program's behaviour. Execution probability can be estimated by simply running the program, and seeing how often each location is executed. To estimate the probability that faults will cause data state errors, sensitivity analysis must simulate various faults, and in order to estimate the probability that data-state errors will affect the program's output, sensitivity analysis must simulate different data-state errors. An empirical evaluation of this idea is lacking in the paper. However, the described sequence of actions has become the standard experimental procedure adopted by existing works dealing with FEP.

The later work by Voas and Miller~\cite{VoasMiller1994} views each location in the program as a point where an assertion checking the internal state can be placed. The authors advocate a middle ground between no assertions at all (the most common practice) and the theoretical ideal of assertions at every location, introducing the problem of \textit{optimal oracle placement}. The authors propose locations determined by sensitivity analysis for the assertion placement. This idea was used in later works for determining the optimal data set for output-based oracles~\cite{StaatsGH12} and for determining locations to place input-specific internal oracles~\cite{XiongH0ZZL15}.

\subsection{FEP in Different Studies}

A number of studies provide evidence of the occurrence of FEP. The motivations for these studies vary. Some of them analyse specific cases such as propagation of error codes in file systems~\cite{Gunawi2008,Rubio-Gonzalez2009PLDI,Rubio-Gonzalez2009SigPlan}. However, the majority are motivated by FEP being undesirable for \textit{Coverage-Based Fault Localization} (CBFL) techniques~\cite{Wong2007, Baudry2006,Jones2004}. Therefore, there is a large body of work aiming to reduce the vulnerability of CBFL to FEP and this usually includes measuring the prevalence of FEP in the subjects  of the experiments~\cite{masri2009empirical,masri2010,MasriA14,miao2012,li2012using,XuePN14,zheng2013}.  The work by Xiong et al.~\cite{XiongH0ZZL15} analyses FEP to demonstrate that faults cannot always be captured by traditional output-based oracles and oracles defined on internal states of the software can further improve the power of testing systems. Androutsopoulos et al.~\cite{AndroutsopoulosCDHH14} introduce an information theoretic formulation of FEP based on measures of conditional entropy, as well as 5 metrics that are well-correlated with FEP.

However, all of these studies use different terms (\textit{error masking}, \textit{fault masking}, \textit{strong/weak coincidental correctness})  and definitions to express closely related notions, or use the same term with different meanings. To make existing studies comparable, we unify existing terms and definitions as follows, using the PIE framework as a well understood basis for unification:

\begin{itemize}  
\item \textit{Coincidental Correctness} (CC) occurs when a fault is \textit{executed}, but is not \textit{propagated} to the output.
\item \textit{Failed Error Propagation} (FEP) occurs when a fault is \textit{executed}, it \textit{infects} the data state, but it does not \textit{propagate} to the output.
\end{itemize}
 
Table  \ref{tab:fep_studies} provides an overall summary of the studies on FEP. 
%Column \textit{Language} shows the programming language of the subject programs. Column \textit{Fault Type} shows what type of faults were analysed in the corresponding study: synthetic mutations, faults seeded into the source code by developers or real faults. Column \textit{Number of Faults} shows how many faults of the given type were generated. \textit{FEP type} shows whether the study measured Coincidental Correctness or Failed Error Propagation, and  \textit{FEP ratio} shows the average rate of CC/FEP reported in the study. 
As we can see from the table, previous work  tends to suggest that there is a nontrivial proportion of faults that are subject to FEP. However, the ratio of FEP varies across different studies substantially: from 7.4\% to 78\%. The majority of these studies come from \textit{fault localisation} and 4 out of 8 studies in the table use subjects from the same Siemens Suite. The majority of the studies use mutants or seeded faults, which are used to simulate real faults. Two studies ~\cite{DaranT96, WangCCZ09} analysing real faults use a single C subject and considering respectively 12 and 38 real faults for it. Only two previous papers considered FEP for Java programs. However, neither of them attempted to measure FEP on real faults. 

%There are studies 

\section{Experimental Procedure} \label{sec:proc}

In this section, we provide the details of the procedure followed in this paper to measure FEP on real faults. We first present the benchmark used in the empirical study. To obtain statistically significant measurements we needed large pools of inputs exercising the faulty statements. We describe the automated test case generation approach adopted this purpose. Then, we describe how execution traces of faulty and fixed programs have been aligned so as to compute state differences at corresponding program points. Finally, we give detailed information on how the FEP measures were obtained from the aligned traces.

\subsection{Benchmark}
To analyse FEP in programs with real faults we used Defects4J \cite{just:fse}\cite{just:defects4j}, a large scale database of existing faults, which contains 395 real bugs from 6 real-world Java open source projects (442 classes and 71455 SLOC per project on average). 
%Indeed, Defects4J has become the standard, reference benchmark for the evaluation of testing techniques on real faults~\cite{Just:2014:MVS,Ma:2015:GRT,B.Le:2016:LBF,Lu:2016:RTP,Mart:2016:APR,Xin:2017:ITP,Pea:2017:EIF,Xiong:2017:PCS,Sham:2015:AUT}.

For each bug we identify whether it is suitable for our study by checking if its fix is a change in a method/constructor. The results show that for 9 out of 395 bugs, the fix is a change in other class members as instance/static variables, static initialisation blocks or in the class declaration itself (as the interfaces it implements). We exclude these bugs from our study. For system-level FEP analysis we can use only bugs from projects with system-level functionality. This condition is satisfied for 132 bugs from the Closure Compiler project, as the remaining 5 projects in Defects4J are libraries. For unit-level FEP analysis we exclude methods/constructors with only one statement, as there is no possibility for internal or external FEP in them. 

Table~\ref{tab:defects4j_info} shows the projects contained in Defects4J, the number of bugs and changed methods/constructors for each of them. In total, we have 386 bugs and 459 methods/constructors available for unit-level FEP analysis and 132 bugs available for system-level FEP analysis.

\HIDE{total number of all available bugs and bugs with fix in methods/constructor; a total number of methods/constructors changed as a result of bug fix and the ones with more than 1 LOC.}

\begin{table}[htb]
\caption{Defects4J Projects (M/C means Methods/Constructors)}
\label{tab:defects4j_info}
\begin{small}
\begin{center}
\begin{tabular}{|l|l|r|r|r|r|}
\hline
\multicolumn{1}{|l|}{\textbf{Project Name}} &
\multicolumn{2}{|c|}{\textbf{Bugs}} &  
\multicolumn{2}{|c|}{\textbf{Number of M/C}} \\
\hline
  &  \textbf{Fix in M/C} & \textbf{All}  & \textbf{> 1 LOC} &\textbf{All} \\
JFreeChart & 25 & 26 & 36 & 39 \\
Closure Compiler &  132 & 133 & 153 & 172 \\
Commons Lang & 62 & 65 & 73 & 84\\
Commons Math & 104 &106 & 126 & 146\\
Mockito & 37 & 38 & 31 & 68\\
Joda Time & 26 & 27 & 40 & 51\\
\hline
Total & 386 & 395 & 459 & 560\\
\hline
\end{tabular}
\end{center}
\end{small}
\end{table}

\subsection{Input Generation}
As FEP might occur only for specific inputs, to measure its probability we need a large number of executions that cover the faulty statements. 

To obtain these executions for unit-level FEP analysis, we extended the EvoSuite~\cite{fraser:evosuite} test case generator. We identify the difference between the buggy and fixed versions of the method in terms of lines of code. The standard line coverage criteria of EvoSuite aims at generating a test suite that covers all lines of code. However, we need to cover only lines of code that contain faulty statements. Moreover, we need these lines of code to be covered multiple times, by different test cases. For this purpose, we made changes to EvoSuite's implementation, so as to handle the following new parameters: 

\begin{enumerate}
\item \textit{line\_list}: list of lines of code to be covered by the generated test cases;
\item \textit{goals\_multiply}: number of times each line should be covered.
\end{enumerate}

In our experiments we aimed to have 1,000 different executions covering each fault. This number of executions was judged a good balance between the total time spent on each experiment and the resulting size of the test pool per bug. Achieving a coverage goal results in a single execution that covers a single line. Each bug consists of a number of lines, the lines in \textit{line\_list}. Therefore, we calculate the value for the parameter \textit{goals\_multiply} by dividing 1,000 by the size of the list provided as the \textit{line\_list} parameter and round this number. We run our extension of EvoSuite giving it a cumulative, maximum search budget of 10,000 seconds (i.e., a maximum of 10 seconds per coverage goal). 
\HIDE{
\david{Do we need the following sentence?} Since we generate 1,000 test cases per bug and these tests are generated on the faulty program version (as a developer would do to expose faults during development), each bug requires a separate test generation process, executed on a distinct program version, i.e., the one containing the considered bug.
}

For system-level FEP analysis, we needed inputs for \textit{Closure Compiler}, which is a tool that accepts a JavaScript file as an input, analyzes it, removes dead code and rewrites and minimizes what's left. We downloaded the 15 most highly trending JavaScript projects from GitHub\footnote{https://github.com/trending/javascript, downloaded on 18.09.2017}, containing 3779 JavaScript files in total, and used these files as inputs to our system.
 
\subsection{Trace Alignment}

To identify the cases of internal and external FEP, we trace both faulty and fixed methods, and we compare the values of variables at corresponding program points in the faulty and fixed versions of the method. 
In simple scenarios, where the fault fix requires only a change in an existing statement, the correspondence between program points is trivially by position in the linearly ordered sequence of statements, i.e., corresponding program points are program points with the same line number. However, in more complex cases, in which the fix requires the addition of new statements and/or the deletion of existing statements, the statement sequences aligned by order must exclude program points that refer to added/deleted statements. Hence, the identification of the corresponding statements can be obtained by calculating the tree edit distance between the Abstract Syntax Trees (AST) of faulty and fixed methods. The \textit{tree edit distance}~\cite{ShashaZhang} is the minimal-cost sequence of node edit operations that transform one tree into another, where the allowed edit operations are: CHANGE, INSERT, DELETE. 

We represent the source code of faulty and fixed versions of a method as an AST using JavaParser\footnote{http://www.javaparser.org}.  We adapted the tree edit distance computation algorithm described in~\cite{ShashaZhang} so that it works with nodes which are objects of JavaParser's \verb"Node" type. We assign the cost of 1 to the three edit operations (CHANGE, INSERT and DELETE) supported by the algorithm. As a result, we get an edit sequence which converts one tree into another and therefore a faulty method into the fixed one. 

\begin{figure}[htb]
\begin{tabular}{cc}
\begin{lstlisting}
1    public int test(int x) {
2       //pp0 (b, c, d)
3   	  int y = x + 1;
4       //pp1 (b, c)
5       y = y % 4;
6       //pp2 (b, c) //pp1 (d)
7       return y; 
8    }
\end{lstlisting} & \\
\vspace{-0.3cm} \\
\multicolumn{2}{c}{(a)} \\
\vspace{0.05cm} \\
\begin{lstlisting}
1    public int test(int x) {
2       //pp0
3   	  int y = x + 1;
4       //pp1
5       y = y % 3;
6       //pp2
7       return y; 
8    }
\end{lstlisting} & 
\begin{minipage}{4cm}
\begin{small}
\begin{verbatim}
KEEP int y = x + 1;
CHANGE y = y % 4; to 
       y = y % 3;
KEEP return y;
\end{verbatim}
\end{small}
\end{minipage} \\
\vspace{-0.3cm} \\
\multicolumn{2}{c}{(b)} \\
\vspace{0.05cm} \\
\begin{lstlisting}
1    public int test(int x) {
2       //pp0
3   	  int y = x + 1;
4	     y = y * 3;
5       //pp1
6       y = y % 4;
7       //pp2
8       return y; 
9   }
\end{lstlisting} & 
\begin{minipage}{4cm}
\begin{small}
\begin{verbatim}
KEEP int y = x + 1;
INSERT y = y * 3;
KEEP y = y % 4;
KEEP return y;
\end{verbatim}
\end{small}
\end{minipage} \\
\vspace{-0.3cm} \\
\multicolumn{2}{c}{(c)} \\
\vspace{0.05cm} \\
\begin{lstlisting}
1    public int test(int x) {
2        //pp0
3        int y = x + 1;
4        //pp1
5        return y; 
6   }
\end{lstlisting} & 
\begin{minipage}{4cm}
\begin{small}
\begin{verbatim}
KEEP int y = x + 1;
DELETE y = y % 4;
KEEP return y;
\end{verbatim}
\end{small}
\end{minipage} \\
\vspace{-0.3cm} \\
\multicolumn{2}{c}{(d)} \\
\end{tabular}
\caption{Buggy method (a) and hypothetical fixed versions, obtained by changing an existing statement (b), by adding a new statement (c), or by removing an existing statement (d)}
\label{fig:method_fixes}
\end{figure}

Figure~\ref{fig:method_fixes} (a) shows an example of a simple method \texttt{test(int x)}, which, for the purpose of the explanation, we consider as a buggy method. Three hypothetical fixes are shown in Figure~\ref{fig:method_fixes} (b), (c), (d), involving respectively the change of an existing statement, the addition of a new statement and the deletion of an existing statement. The edit scripts automatically produced by our implementation of the tree edit distance algorithm are shown in the right column of the figure.

\begin{algorithm}[htb]
\DontPrintSemicolon
\SetKwFunction{visit}{visit}
\SetKwProg{myproc}{Procedure}{}{}
\myproc{\visit{n, i}}{
\KwIn{\;
$n$: AST node to be visited\;
$i$: instrumentation index\;
}
\Begin{
	\If{$n$ is labeled as KEEP or CHANGE $\wedge$ type($n$) is not (RETURN or THROW)} {
		\While{next($n$) is labeled as DELETE or INSERT} {
		    $n$ := next($n$) \;
		}
		$i$ := $i$ + 1 \;
		instrumentAfter($n$, $pp_i$) \; 
	} \Else {
		visit(next($n$), $i$) \;
	}
	\If{type($n$) is not (FOR or WHILE)} {
	\For{$m \in $ children($n$)} {
		visit($m$, $i$) \;
	}
	}
}
}
\caption{Program point instrumentation}\label{algo:visit}
\end{algorithm}

After the edit script is generated, we start the instrumentation process. For both the buggy and fixed versions of the method we instrument the starting program point \textit{pp0}. Then we visit the nodes in the ASTs of the two methods, according to the pseudocode of Algorithm \ref{algo:visit}.  If a node is associated with the KEEP or CHANGE operators and if it is not of the RETURN or THROW statement types, we instrument the program point after this node (line 7), skipping any sequence of INSERT and DELETE nodes (lines 4-5). Otherwise, if it is associated with the INSERT or DELETE operators, we skip the program point and proceed with the next node (line 9).  Then, if the node is not a \textit{while} or \textit{for} loop, the visit proceeds recursively on the subtrees (lines 10-12). We exclude program points within loops because of the practical difficulty of defining oracles for the program state inside a loop.

By following this procedure, we obtain the program point correspondence indicated within comments in Figure~\ref{fig:method_fixes}, associating program points in version (a) with those in (b), (c), (d). The placement of program points in $\langle$(a), (b)$\rangle$ is straightforward. In case of $\langle$(a), (c)$\rangle$, the program point before added statement \verb+y = y * 3+ is skipped during the visit of (c) due to the while loop at lines 4-5 in Algorithm~\ref{algo:visit}. Similarly, the deleted statement \verb+y = y % 4+ 
is jumped over during the visit of (a). As a consequence, the program point after \verb"y = x + 1" 
in (a) has no corresponding program point in (d).

\subsection{Measuring FEP}

\begin{algorithm}[htb]
\DontPrintSemicolon
\KwIn{\;
$type$ = $\langle sys$  $\mid$ $unit \rangle$: type of analysis, system-level or unit-level\;
$out$, $out'$: output of the system, used only for system-level analysis\;
$ext$, $ext'$: externally observable state after buggy/fixed methods have been executed\;
$pp= \langle pp_0, \dots, pp_n \rangle$: program points executed in fixed method\;
$pp' = \langle pp'_0, \dots, pp'_k \rangle$: program points executed in buggy method\;
$s$, $s'$: state by program point in buggy/fixed methods\; 
}
\KwResult{\;
\textit{fepType}:  $\langle$ sysFEP $\mid$ intFEP $\mid$ extFEP $\mid$ noFEP $\rangle$\;
}

\Begin{

        \If{$type$ = $unit$  $\&\&$ $ext$ $\neq$ $ext'$} {
		\Return noFEP \; 
	}
	
	\If{$type$ = $sys$}{
        		\If{$out$ $\neq$ $out'$} {
			\Return noFEP \; 
		}	
		
	        \If{$ext$ $\neq$ $ext'$} {
		\Return closure(sysFEP, $type$) \; 
		}
        }
	
	\If{$s[pp_n] \neq s'[pp'_k]$} {
		\Return closure(extFEP, $type$) \;
	}
	
	\If{$pp \neq pp'$} {
		\Return closure(intFEP, $type$) \;
	}
	
	\For{$i \in [1:n-1]$} {
		\If{$s[pp_i] \neq s'[pp'_i]$} {
			\Return closure(intFEP, $type$) \;
		}
	}
	\Return noFEP \;
}
\nonl where \textit{closure(FEP, $type$)} applies the implication \textit{intFEP} $\Rightarrow$ \textit{extFEP} $\Rightarrow$ \textit{sysFEP} when $type$ = $sys$ and \textit{intFEP} $\Rightarrow$ \textit{extFEP} when $type$ = $unit$.
\caption{Measuring FEP}\label{algo:measuring-fep}
\end{algorithm}

After running the generated inputs on the instrumented methods, we obtain the values of variables at each program point, for each execution. Algorithm \ref{algo:measuring-fep} shows how we identify whether a given execution is subject to FEP. 

As shown at lines 2-3, if unit-level FEP analysis is performed and the externally observable state is affected by the faulty execution as compared to the fixed execution, we report no FEP. Similarly, if system-level FEP analysis is performed and the output of the system is affected by the fault, we report no FEP (lines 5-6). However, if the output of the system remains the same for the faulty and fixed execution, but the externally visible state is different, we report system-level FEP (lines 7-8). Otherwise, we check whether the state at the program point before return is different (lines 9-10) and if it is so, we report external FEP. If there is no external FEP, and the program points traversed by the executions in the buggy and fixed methods are different (lines 11-12), then internal FEP is detected -- here, the executions in the faulty and the fixed methods took different paths, so if we place an internal oracle in the buggy method checking for the predicates in the path of the fixed execution, it would detect the fault. In the case that all program points in the two executions are the same, we iterate through them and report internal FEP if the state in at least one aligned pair of them is different (lines 13-15). Finally, it is also possible that for some inputs, the bug in the method does not lead to any changes at all in the pair of executions being compared. This is also a case of no FEP (line 16). 
The FEP value returned by the algorithm is expanded with the addition of the subsumed values (invocation of \textit{closure} in Algorithm~\ref{algo:measuring-fep}). This means for instance that if intFEP is reported for a system level analysis, extFEP and sysFEP are also reported as true.

We run this algorithm for each system or method/constructor execution, as appropriate, and then we calculate the proportion of either system-level FEP or unit-level FEP (both internal and external) across all of the executions that cover the considered fault, to estimate the probability of FEP for such a fault. 
In this algorithm, the value of variables at each program point may represent Java objects that need to be stored and compared with each other. For this we use the XStream framework\footnote{http://x-stream.github.io/}, which can serialize any Java object without requiring their classes to implement the\- \textit{java.io.Serializable} interface (including private and final fields). We serialize these objects to JSON format and consider two objects equal when their JSON representations are the same.

\section{Results}  \label{sec:results}

\begin{table*}[htb]
\caption{Internal and external FEP on real faults (RQ1)}
\label{tab:defects4j_fep_real}
\begin{small}
\begin{center}
\begin{tabular}{|l|r|r|r|r|r|r|}
\hline
\textbf{Project} & \textbf{Changed} & \textbf{Methods} & \textbf{Number of} & \textbf{Externally}  & \textbf{Int} & \textbf{Ext} \\
\textbf{Name} & \textbf{Methods} & \textbf{with TS} & \textbf{Executions} & \textbf{Detectable}  & \textbf{FEP} & \textbf{FEP} \\
\hline
JFreeChart & 36 & 28 & 18,785 & 10,678 & 0 & 0 \\
Closure Compiler & 153 & 102 & 89,078 & 42,078 & 0 & 0\\
Commons Lang & 73 & 54 & 35,153 & 24,854 & 0 & 0\\
Commons Math & 126 & 92 & 78,489 & 45,065 & 0 & 0\\
Mockito & 31 & 25 & 20,967 & 8,348 & 0 & 0\\
Joda Time & 40 & 28 & 15,900 & 7,987 & 0 & 0\\
\hline
Total & 459 & 329 & 258,372 & 139,010 & 0 & 0\\
\hline
\end{tabular}
\end{center}
\end{small}
\end{table*}

\subsection{Research Questions}

We have conducted a set of experiments to answer the following research questions:

\begin{itemize}
  \item \textbf{RQ1}: \textit{What is the prevalence of unit-level failed error propagation with real faults?}
  \item \textbf{RQ2}: \textit{Does the prevalence of unit-level failed error propagation change if real faults are replaced by mutants?}
  \HIDE{\item \textbf{RQ3}: \textit{Do existing, manual test cases include assertions capable of detecting all faults detectable by an external oracle?}}
  \item \textbf{RQ3}: \textit{Does the prevalence of failed error propagation with real faults change if it is measured at the system level instead of unit level?}
  %\item \textbf{RQ4}: \textit{What are the factors affecting the existence or absence of FEP?}
\end{itemize}

RQ1 is the key research question that motivates this study. The answer to this question has implications for oracle placement. It is potentially relevant for both practitioners and researchers, since it estimates the probability of missing / detecting a fault depending on where oracles are placed (i.e., internally, at return points, or externally).

While, to the best of our knowledge, no previous study investigated the occurrence of FEP on real Java faults, there are experimental results~\cite{AndroutsopoulosCDHH14,MasriA14,XiongH0ZZL15} on FEP computed when mutations are used as surrogates for real faults in both Java and C. Such results provide evidence for the occurrence of FEP on mutants. With RQ2 we want to investigate whether results on mutants correspond to the results obtained on real faults.

Since a system level execution typically involves a long chain of concatenated unit level executions, there is potentially more opportunity for a corrupted state to disappear during such a system-level execution, becoming undetectable at the output. In RQ3 we want to check whether the prevalence of FEP changes (and in particular, whether it increases) when we consider test executions at the system level instead of unit level as expected. 

We also report some observations obtained from a qualitative analysis performed to better understand the patterns of prevalence behind FEP or no FEP, either with real faults or with mutations, considering the root cause of each occurrence.

%%%%%%% what follows is the old presentation of RQ3 %%%%%%%%%%%%%
%In the absence of FEP, external oracles could in principle detect faults as soon as such faults produce an execution state that diverges from the expected one. However, external oracles may or may not be strong enough to detect all deviations from the expected behaviour that they could indeed detect, depending on how they are written in practice. This is investigated in RQ3. Inner and/or return point oracles would be still useful, even in the absence of external FEP, when the external oracles that developers write in practice are not actually capable of capturing all possible deviations from the expected behaviours. Alternatively, in the presence of weak oracles, oracle improvements techniques~\cite{JahangirovaCHT16} can be used to strengthen the external oracles that miss some externally visible deviations from the intended program behaviour.

\subsection{Experimental Data}

We have made all experimental data available in a package that can be downloaded from http://anonymized-for-double-blind-review.

\subsubsection{RQ1 (FEP in programs with real faults)}
 
Table \ref{tab:defects4j_fep_real}  shows a summary of the results obtained in our experiments. Column \textit{Changed Methods} indicates the overall number of methods changed as a result of a bug fix, while column \textit{Methods with TS} reports the number of methods for which our extended version of EvoSuite was able to generate a large test suite, consisting of test cases that exercise the faulty statements. While the target size of these test suites was 1,000 test cases, sometimes EvoSuite generated slightly smaller test suites in the allowed generation time (the average test suite size is 863).

Column \textit{Number of Executions} shows the overall number of executions obtained as a result of running the test cases. Column \textit{Externally Detectable} shows the number of executions where the fault resulted in a program state deviation that is observable outside of the methods. Columns \textit{Internal FEP} and \textit{External FEP} show that, among the 258,372 executions, the fault was externally observable in 139,010 cases (53.8\%). In the remaining cases (119,362 test case executions), in order for FEP to happen an internal program state deviation, not propagated to the output, should be observed. However, this was never the case. There was no single case where an internal state deviation occurred, i.e. no state infection.

We have tested the statistical significance of our results, which depends on sample size and observed values.
According to the Pearson-Klopper method for calculating binomial confidence intervals, internal/external FEP is in the range [0:$1.43^{-5}$] with mean = 0 at confidence level 95\%. This means that even if intFEP and extFEP could occur in other subjects (we might have not observed it just by chance), their likelihood can be assumed to be very low with high confidence.

\begin{center}
\fbox{
\begin{minipage}[t]{0.95\linewidth}
\textbf{RQ1}: \textit{Our experiments show that the probability of unit-level FEP in Java methods with real faults is extremely low.}
\end{minipage}
}
\end{center}
 
\subsubsection{RQ2 (FEP in mutated programs)}

\begin{table}[htb]
\caption{Methods from benchmark grouped by LOC}
\label{tab:defects4j_method_loc}
\begin{small}
\begin{center}
\begin{tabular}{|l|r|r|r|r|r|r|}
\hline
\textbf{Project Name} & \textbf{2-25} & \textbf{26-50} & \textbf{51-} & \textbf{101-} & \textbf{>200}\\
\textbf{} & \textbf{ } & \textbf{ } & \textbf{100} & \textbf{200} & \textbf{}\\
\hline
JFreeChart & 22 & 6 & 5 & 3 & 0 \\
Closure Compiler & 60 & 45 & 35 & 8 & 5 \\
Commons Lang & 34 & 16 & 11 & 12 & 0\\
Commons Math & 59 & 22 & 24 & 16 & 5\\
Mockito & 25 & 6 & 0 & 0 & 0\\
Joda Time & 26 & 12 & 2 & 0 & 0\\
\hline
\end{tabular}
\end{center}
\end{small}
\end{table}

For RQ2, instead of real faults we consider faulty versions of methods obtained by means of mutation analysis (i.e., we generate mutants of the fixed Defects4J methods). As we need a large test suite for each mutant and the number of mutations generated per method can be high, we did not conduct this analysis on all the methods available in the benchmark. Instead, we sampled the methods based on their lines of code, we divided methods into 5 groups: 2-25 LOC, 26-50 LOC, 51-100 LOC, 101-200LOC, > 200LOC. Table \ref{tab:defects4j_method_loc} shows the number of methods in each group for each project. We  randomly selected one method from each group for each project and we generated mutants for the selected representative using Major\cite{just:major} and applying all the mutation operators available in this tool. Then, among the generated mutations, we selected only strongly killable mutants, to avoid the inclusion of equivalent mutants. In fact, an internal state deviation in an equivalent mutant is always associated with external FEP, but this is by definition a false positive, because the internally observed difference is not an indicator of a fault: since the mutant is equivalent to the original program, it does not introduce any fault into the program, so there is no fault to be detected internally at all. Hence, we conservatively measure FEP only on mutants proved to be strongly killable by test generation. In cases when EvoSuite was unable to generate a large test suite for any of the mutations of a method, or when none of them is strongly killable, we randomly select another method from the group. 

\begin{table}[htb]
\caption{Mutants generated.}
\label{tab:defects4j_stk_mutations}
\begin{small}
\begin{center}
\begin{tabular}{|l|r|r|r|r|r|r|}
\hline
\textbf{Project Name} & \textbf{Mutants} & \textbf{Strongly} \\
\textbf{} & \textbf{} & \textbf{Killed} \\
\hline
JFreeChart & 37 & 25 \\
Closure Compiler & 468 & 350 \\
Commons Lang & 360 & 215\\
Commons Math  & 765 & 502 \\
Mockito & 28 & 15  \\
Joda Time & 212 & 18 \\
\hline
Total & 1870 & 1125 \\
\hline
\end{tabular}
\end{center}
\end{small}
\end{table}

\begin{table}[htb]
\caption{Internal/external FEP on mutants (RQ2)}
\label{tab:defects4j_fep_mutations}
\begin{small}
\begin{center}
\begin{tabular}{|l|r|r|r|r|r|r|}
\hline
\textbf{Project} & \textbf{Num of} & \textbf{Externally}  & \textbf{Int} & \textbf{Ext} \\
\textbf{Name} & \textbf{Execs} & \textbf{Detectable}  & \textbf{FEP} & \textbf{FEP} \\
\hline
JFreeChart & 25,842 & 6,217 & 0 & 0 \\
Closure Compiler & 320,678 & 180,562 & 2,587 & 4,783\\
Commons Lang & 89,043 & 45,800 & 1,567 & 2,623 \\
Commons Math & 422,068 & 200,865 & 10,222 & 25,956\\
Mockito & 16,284 & 7,321 & 0 & 0\\
Joda Time & 15,460 & 8,970 & 0 & 0\\
\hline
Total & 889,375 & 449,735 & 14,376 & 33,362 \\
\hline
\end{tabular}
\end{center}
\end{small}
\end{table}

Table \ref{tab:defects4j_stk_mutations} shows the overall number of mutants and the number of mutants that are strongly killable by the generated test suites. We can see from Table \ref{tab:defects4j_fep_mutations} that when we replace real faults with mutations, for 3 subjects there are cases of both internal and external FEP.  Among all  831,789 executions in these 3 subjects, 51\% of  faults were externally detectable. In 1.6\% of executions there was an occurrence of internal and in 3.7\% of external FEP. In the remaining cases (46.9\%) the internal state was always identical to the expected one, i.e., the fault did not infect the execution.

According to the Pearson-Klopper method, internal FEP is in the range [0.0159:0.0164], with mean = 0.0161, at confidence level 95\%; external FEP is in the range [0.0210:0.0216], with mean = 0.0213, at confidence level 95\%.

\begin{center}
\fbox{
\begin{minipage}[t]{0.95\linewidth}
\textbf{RQ2}: \textit{Mutants behave in a substantially different way than real faults when FEP is considered for Java methods: there is higher probability of both internal and external FEP when the fault is introduced by mutation.}
\end{minipage}
}
\end{center}

\subsubsection{RQ3 (System-level FEP)}

\begin{table}[htb]
\caption{System-Level FEP on real faults (RQ3)}
\label{tab:system_fep_real}
\begin{small}
\begin{center}
\begin{tabular}{|c|r|r|r|r|r|r|}
\hline
\textbf{Closure} &  \textbf{Num of}  & \textbf{Externally} & \textbf{Sys}  & \textbf{Int} & \textbf{Ext} \\
\textbf{Bug ID} & \textbf{Execs} & \textbf{Detectable}   & \textbf{FEP}  & \textbf{FEP} & \textbf{FEP} \\
\hline
\textbf{1} & 20 & 12 & 8 & 8 & 8  \\
\textbf{4} & 15 & 5 & 8 & 0 & 0\\
\textbf{8} & 200 & 159 & 41  & 0 & 0\\
\textbf{13} & 36 & 24 & 0 & 0 & 0\\
\textbf{16} & 15 & 10 & 0 & 0 & 0\\
\textbf{20} & 22 & 22 & 0 & 0 & 0\\

\textbf{21} & 4 & 4 & 0 & 0 & 0  \\
\textbf{22} & 4 & 4 & 0 & 0 & 0\\
\textbf{29} & 1 & 1 & 0  & 0 & 0\\
\textbf{34} & 13 & 10 & 0 & 0 & 0\\
\textbf{50} & 1 & 1 & 0 & 0 & 0\\
\textbf{52} & 13 & 13 & 0 & 0 & 0\\

\textbf{56} & 2 & 2 & 0 & 0 & 0  \\
\textbf{60} & 5 & 5 & 0 & 0 & 0\\
\textbf{62} & 57 & 50 & 0  & 0 & 0\\
\textbf{63} & 57 & 50 & 0 & 0 & 0\\
\textbf{87} & 23 & 12 & 3 & 0 & 0\\
\textbf{115} & 3 & 3 & 0 & 0 & 0\\

\textbf{116} & 4 & 4 & 0 & 0 & 0  \\
\textbf{127} & 14 & 14 & 0 & 0 & 0\\
\textbf{131} & 9 & 9 & 0  & 0 & 0\\
\textbf{133} & 10 & 10 & 0 & 0 & 0\\
\hline
Total & 528 & 424 & 60 & 8 & 8 \\
\hline
\end{tabular}
\end{center}
\end{small}
\end{table}

For RQ3 we have run \textit{Closure Compiler} on 5,070 different JavaScript input files, on 132 bugs of this project. For each bug, we made a run on both faulty and fixed versions of the system and saved the pairs of outputs and method executions obtained. The output of \textit{Closure Compiler} is also a JavaScript file and if the output files generated are different we consider the execution to be \textit{Externally Detectable}. Table \ref{tab:system_fep_real} lists the ID of the bugs which we were able to execute with our inputs. 22 bugs out of 132 were executed leading to an overall number of 528 executions. For each of these 22 bugs there was at least one execution which was externally detectable, i.e. that caused a change in the output file generated by \textit{Closure Compiler}.  Overall, 424 out of 528  (80.3\%) executions were externally detectable. 60 executions (11.4\%) provide evidence of FEP in 4 different bugs. For 8 executions (1.5\% of all executions) of \textbf{Bug 1} we observed unit-level internal and external FEP.  This bug affects neither the externally observable state of the class nor the final output of the system. However, it causes the program states in faulty and fixed versions to differ, which is evidence of both Internal and External FEP. During the unit-level analysis our test case generator was not able to generate any test cases for this bug, therefore no unit-level FEP was reported in RQ1.

According to the Pearson-Klopper method for calculating binomial confidence intervals, internal FEP is in the range [0.0878:0.1438], with mean = 0.1136, at confidence level 95\%.

\begin{center}
\fbox{
\begin{minipage}[t]{0.95\linewidth}
\textbf{RQ3}: \textit{The prevalence of FEP changes when we test programs at the system level instead of unit level: 11.4\% of the overall executions provide evidence of system-level FEP, which is substantially higher than the probability of unit-level FEP, both with real faults and with mutants.}
\end{minipage}
}
\end{center}

\subsubsection{Qualitative analysis of factors affecting FEP} 

To understand the reasons behind the absence of FEP in unit level programs with real faults and their existence in the mutations of these, we performed a qualitative analysis on all of the 386 bugs from the Defects4J benchmark and all of the mutations generated by Major. For methods from Defects4J, we manually compared the buggy version of the methods with the fixed version and analysed the bug fixes. As a result of this analysis we identified two major classes of explanations for the absence of FEP: (1) the fix of the bug affects the output directly; (2) the state change resulting from the fix is such that it must necessarily propagate to the output. In case (1), clearly both internal and external FEP are impossible, since all state deviations are immediately returned to the external oracle. In case (2), the state change propagates to the output because the computation performed between the fault and the return statement does not ``squeeze'' the state \cite{AndroutsopoulosCDHH14}.
% (i.e., it never collapses correct and incorrect values into the same value, as happens e.g. with statement \verb+x = x % 4+ in Figure~\ref{fig:example}).

\paragraph{Bug fix affects the output directly}

\HIDE{
\begin{figure}[htb]
\begin{lstlisting}
1  public Complex divide(double divisor) {
2	  if (isNaN || Double.isNaN(divisor)) {		        
3         return NaN;		            
4     }		        
5     if (divisor == 0d) {		       		            
6         return NaN;	
7         //return isZero ? NaN : INF; 	           
8     }		        
9     if (Double.isInfinite(divisor)) {		       
10         return !isInfinite() ? ZERO : NaN;		            
11     }		        
12    return createComplex(real / divisor,		      
13                             imaginary  / divisor);		                            
14  }
\end{lstlisting}
\vspace{-0.5cm}
\caption{Commons Math Bug 46}
\label{fig:math_46}
\end{figure}
}

\begin{figure}[htb]
\begin{lstlisting}
1  public static LocalDate fromDateFields(Date date) {		   
2    if (date == null) 	   
3        throw new IllegalArgumentException
4                   ("The date must not be null");		           
5    // if (date.getTime() < 0) {
6    //    GregorianCal cal = new GregorianCal();
7    //    cal.setTime(date);
8    //    return fromCalendarFields(cal);
9    // }
10    return new LocalDate(		        
11       date.getYear() + 1900,		           
12       date.getMonth() + 1,		            
13       date.getDate());		        
14   }
\end{lstlisting}
\vspace{-0.5cm}
\caption{Joda Time Bug 12}
\label{fig:time_12}
\end{figure}

During manual analysis, one commonly occurring fix pattern was a change in the \textit{return} statement of a method. As fix changes
the \textit{return} statement directly, it is not possible to observe any difference between the fixed and buggy versions at some internal point in the method, so no FEP can be observed in such cases.
%For example, in Figure \ref{fig:math_46} the bug is at line 6 and the fix is as indicated within a comment at line 7. As this fix changes the return statement directly, it is not possible to observe any difference between the fixed and buggy versions at some internal point in the method, so no FEP can be observed in such cases. 
A variant of this pattern is the addition of an \textit{if} statement containing \textit{return} or \textit{throw} statements inside. In Figure \ref{fig:time_12} the bug is fixed by adding the \textit{if} statement at lines 5-9. So whenever this \textit{if} statement is executed, the method will return the object produced by the invocation at line 8. If this differs from the object generated by the faulty version at line 10, the difference will be definitely observable by an external oracle. If it does not differ, we have coincidental correctness, but no FEP.

\begin{table}[htb]
\caption{Fix propagation patterns}
\label{tab:ast_return}
\begin{scriptsize}
\begin{center}
\begin{tabular}{|l|r|r|r|r|r|r|r|r|}
\hline
\textbf{Project} & \textbf{Fixed} & \textbf{Change} & \textbf{Add} & \textbf{Bugs} & \textbf{Prop} \\
\textbf{} & \textbf{meth-} & \texttt{return} & \texttt{if} & & \textbf{to} \\
\textbf{} & \textbf{ods} & \textbf{} & \textbf{} & & \textbf{out} \\
\hline
JFreeChart & 36 & 11 & 3 & 25 & 8 \\
Closure Compiler & 153 & 21 & 13 & 131 & 14 \\
Commons Lang & 73 & 17 & 10 & 61 & 4 \\
Commons Math  & 126 & 28 & 28 & 104 & 17 \\
Mockito & 31 & 9 & 6 & 37 & 1 \\
Joda Time & 40 & 9 & 6 & 26 & 4 \\
\hline
Total & 459 & 95 & 66 & 384 & 48 \\
\hline
\end{tabular}
\end{center}
\end{scriptsize}
\end{table}

To quantify this class of FEP, we considered the edit scripts generated for trace alignment: (1) when the edit script contains a CHANGE operator which changes one return statement into another; or, (2) when the edit script contains an INSERT operator which adds an \textit{if} statement containing a \textit{return} or \textit{throw} statement inside. Table \ref{tab:ast_return} (col. 3-4) reports the number of occurrences of both cases. As we can see, in 35\% of the methods the bug fix includes these type of changes.

\paragraph{Change always propagates to output}

\begin{figure}[htb]
\begin{lstlisting}
1  public double getChiSquare() {		    
2        double chiSquare = 0;		        
3        for (int i = 0; i < rows; ++i) {		       
4            final double residual = residuals[i];		            
5            chiSquare += residual * residual * 
6                             residualsWeights[i];		    
7            //chiSquare += residual * residual / 
8            //             residualsWeights[i];
9        }		        
10      return chiSquare;		      
11   }
\end{lstlisting}
\vspace{-0.5cm}
\caption{Commons Math Bug 65}
\label{fig:math_65}
\end{figure}

Another typical pattern preventing the occurrence of FEP is when a state change resulting from a bug fix always propagates to output. In Figure \ref{fig:math_65} the bug is at lines 5-6 and the fix is as indicated at lines 7-8. If the buggy statement is executed, it might cause a difference in the value of the \textit{chiSquare} variable. However, whenever this happens, this different value is ensured to always propagate to the return statement, hence being externally observable. When there is no difference, we have coincidental correctness, but no FEP.

Table \ref{tab:ast_return} (col. 5-6) shows the number of bugs for each project where this scenario holds. These cases were identified by performing a manual analysis of the fixed bug. Overall, it happens in 13\% of the bugs. %\gunel{the number has dropped a bit after I did analysis on all the subjects}

\section{Implications} \label{sec:impl}

The empirical results presented in this paper have relevant implications for practitioners and researchers:

\paragraph{Internal oracles}

The absence of internal FEP when real faults are considered for Java units (classes) indicates that internal oracles do not have higher fault detection capabilities than return point or external oracles when performing unit testing of classes. Rather than attempting to include assertions about the internal execution state, Java developers might better invest their time to strengthen the assertions that check the program state at return points or within test cases. In fact, if such assertions are sufficiently strong to capture any deviation from the expected execution state, they should miss no fault that manifests itself internally, because the internal state deviation tends to reach them. Researchers interested in Java faults should focus on techniques to improve the oracles that can be defined at return points or within test cases, because these can be made equally effective as internal oracles.

The non-negligible occurrence of FEP at the system level indicates that checking the overall output of a system might be not enough and that probes for the intermediate computations should be inserted into the test case execution to avoid that the effects of faults disappear when proceeding to the computation of the overall system output. While such intermediate oracles can still be based on post conditions or test case assertions, and do not require the observation of internal execution states, they might represent a challenge for system level testing. In fact, at this testing level the system is usually considered as a black box, whose intermediate steps are not visible. According to our results, monitoring and checking such intermediate steps is quite important for avoiding system FEP.

\paragraph{Post-conditions}
The \textit{programming by contract} method prescribes that every method be equipped with pre-conditions, post-conditions and invariants. This approach to programming offers several benefits, among which are the following possibilities: to formally express the specifications that each method must satisfy, in a way that is machine interpretable; to reuse the oracle across test cases; to document a method in an unambiguous way. One may question what part of the execution state should be checked in a post-condition. In fact, at return points the whole internal state of the method under test is accessible. According to our results, the absence of internal FEP indicates that checking the externally visible effects of a method execution is enough to expose faults as soon as they corrupt the execution state. It is unlikely that the effort of creating internal oracles be beneficial to early fault exposure. Rather, postconditions at return points can be focused on the externally visible effects of the execution, disregarding the inner details. This is consistent with the programming by contract paradigm, where only the externally visible contract is typically specified.

\paragraph{Subsystem testing}
The higher prevalence of failed error propagation at system level over unit level might indicate that testing subsystems of the software in isolation could make it easier to expose bugs. While the effect of a bug is externally visible in the class to which it belongs, it is not always visible at the level of the whole system. This might support the idea of bottom-up integration testing, in which we build on unit-level results by testing higher-level combination of units in successively more complex scenarios. 

\paragraph{Mutants vs. real faults}
The software engineering community has witnessed a long debate on the use of mutants as surrogate for real faults~\cite{AndrewsBL05,JustJIEHF14}. Such a replacement may be valid for the purpose of evaluating the adequacy of a test suite, owing to the high correlation between mutation score and fault detection rate. We are interested in investigating the propagation of an error to the oracle that can detect it. Our results show that such propagation is prevalent with mutants, while it is absent with real faults. Our qualitative analysis indicates that mutants corrupt the internal state differently from real faults. In fact, the latter state corruption tends to always generate an externally visible misbehaviour, while the former might remain invisible if only the external state is inspected. Hence, practitioners should not decide where to place their oracles based on the propagation of errors as simulated with mutants. Researchers could investigate mutation operators that behave similarly to real faults with respect to the propagation of the corrupted internal state to the externally visible state.

\paragraph{Previous work}
%%%%%%% Mark's text %%%%%%%%%
As discussed in Section~\ref{sec:impl}, previous work on failed error propagation tended to suggest that there is a nontrivial proportion of faults that manifest this FEP property. Our results differ markedly from these previous findings. One possible explanation could be differences in the subjects and the types of faults. Daran et al.~\cite{DaranT96} analysed 12 real faults in a C program with 1000 lines of code. Wang et al.~\cite{WangCCZ09} analysed 38 real faults in a C program with 6000 lines of code.  By comparison the Defects4J contains 395 real faults which come from six large Java projects. An intriguing possibility lies in the potential differences between the two language (C vs. Java) styles; perhaps some programming languages have inherently higher or lower failed error propagation  propensity than others. Hence, one of the implications of our findings is the pressing need for further work on FEP in different programming languages and corpuses. 
Taken together, our findings and those in the previous literature do tend to suggest that there may be differences between different programming paradigms with respect to error propagation behaviour, and that there are certainly differences between unit and system level FEP. These differences clearly have implications for software testability \cite{binder:design,voas:testability}, because FEP tends to inhibit testability. Such findings may also suggest testability transformations \cite{mhetal:tse-flag,mcminn:co-tetra} that could reduce the likelihood of failed error  propagation, leading to reformulations of software systems (e.g., by inserting probes for intermediate steps) that are inherently more testable.

\subsection{Threats to Validity}

%In this section we discuss potential threats to the validity of our empirical findings. These are mostly in the external and internal validity categories.

Threats to {\em external validity} affect the generalisation of our results. We carried out our experiments on a well established benchmark for Java, Defects4J, which includes 395 real bugs  from 6 different projects. While Defects4J is becoming de-facto a standard benchmark for Java testing, replication of our study on further subjects beyond Defects4J would be quite important. We do not claim generalisability to programming languages other than Java. On the contrary, we suspect that the programming style of Java, which encourages the decomposition of the software into small computations assigned to methods, favours the creation of code units where information is not squeezed when propagating from inner states to the output. Other programming styles might favour the creation of longer and more complex computational units, where information squeezing might be more likely to occur. % \gunel{Maybe we could mention the nature of the bugs in this benchmark, as for each of them there is a test case which can determine the bug, so they all propagate to output for some inputs}

Threats to {\em internal validity} come from factors that could influence our results. Among them, the most important factor that influences our conclusions on the differences between real faults and mutants on FEP, is the set of mutations that have been considered. To limit such a threat, we used a well-established mutation analysis tool, Major. However, different tools and different mutation operators might lead to different sets of synthetic faults. Moreover, we have not been able to perform mutation analysis of all the buggy methods available in Defects4J, because of the enormous computation time involved, since we generate test cases for all mutants that Major produces for each method. We have defined a sampling strategy that takes method size into account, in order to consider representatives of the various possible method size categories. However, this does not ensure that the results obtained on the selected sample would remain exactly the same if extended to the entire dataset of the buggy methods.

\HIDE{
\paolo{The following paragraph can be cut, if we need space}
Another factor that might have influenced the results is the way we filtered equivalent mutants from the full set of mutants generated by Major. We \emph{conservatively} kept only killable mutants. This means that among the excluded mutants, some may be non equivalent and may be subject to FEP. As a consequence, when mutants are considered instead of real faults, our measures of internal/external FEP are conservatively underestimating the true values. Even with such a conservative underestimation, we observed a non negligible number of occurrences. Our conservative underestimation may also explain the lower incidence of FEP on mutants in comparison with the values reported in the literature~\cite{MasriA14,XiongH0ZZL15}.

Finally, our results are potentially affected by the limitations of the test generator used to exercise the faults. EvoSuite was indeed unable to generate large test suites for some faults and EvoSuite might have produced larger test suites if given additional test generation budget. To avoid that small test suites could affect our results, we have excluded all test suite with less than 150 test cases. The test generation budget allocated to EvoSuite (10,000 seconds per test suite) was the maximum compatible with the overall duration of the empirical study.
}

%\gunel{Should we mention test case generation process(more test cases, more budget, another coverage criteria)?}

\section{Conclusion} \label{sec:concl}

In this paper we have presented empirical evidence from a large corpus of real-world faults in Java systems that reveals a surprisingly low unit-level FEP amongst the 386 faults studied.
%These empirical findings contradict earlier work on failed error propagation and, if replicated in other fault corpuses and/or for other languages, would have profound implications for software testing.
On the other hand, with system-level inputs we get a substantially higher rate of FEP. This shows that when oracles are defined for an individual Java class, postconditions that predicate on the externally observable state or test case oracles are sufficient to detect faults as soon as they corrupt the internal state. On the contrary, when we analyse a complete software system, the output alone does not provide enough information to expose faults as soon as they manifest themselves, necessitating the observation of intermediate computation steps. 

When we turn our attention to studying the synthetic faults introduced by program mutants (a widespread practice believed to be good at simulating real faults), we find noticeably different behaviour at the unit level: the artificial faults denoted by mutants {\em do} exhibit substantial FEP, unlike the real faults we studied. 
%These findings concerning mutants provide additional nuances on earlier work on the suitability of mutation testing for simulating real faults.  
While such synthetic faults may be good proxies for estimating whether test cases that reveal them will also reveal real faults, there do appear to be non-trivial differences in the behaviour of synthetic faults and real faults, with respect to their error propagation in Java classes. 

These findings suggest further work to investigate the prevalence of FEP in other programming languages and bug data sets, and the need to further investigate the relationship between mutation testing and real faults. 
We studied only single faults, but future work could also extend our findings to multiple faults, which may have additional implications for higher order mutation testing, one of the main motivations of which is the ability to model fault masking.

\newpage %% (**) so that reference come an a fee page - looks nicer. 
\bibliographystyle{abbrv}
%\bibliography{refs,slice,david,extra-refs}

\begin{thebibliography}{10}

\bibitem{AmmannOffutt2008}
P.~Ammann and J.~Offutt.
\newblock {\em Introduction to Software Testing}.
\newblock Cambridge University Press, New York, NY, USA, 1 edition, 2008.

\bibitem{AndrewsBL05}
J.~H. Andrews, L.~C. Briand, and Y.~Labiche.
\newblock Is mutation an appropriate tool for testing experiments?
\newblock In {\em 27th International Conference on Software Engineering
  (ICSE)}, pages 402--411, 2005.

\bibitem{AndroutsopoulosCDHH14}
K.~Androutsopoulos, D.~Clark, H.~Dan, R.~M. Hierons, and M.~Harman.
\newblock An analysis of the relationship between conditional entropy and
  failed error propagation in software testing.
\newblock In {\em 36th International Conference on Software Engineering,
  {ICSE}}, pages 573--583, 2014.

\bibitem{B.Le:2016:LBF}
T.-D. B.~Le, D.~Lo, C.~Le~Goues, and L.~Grunske.
\newblock A learning-to-rank based fault localization approach using likely
  invariants.
\newblock In {\em Proceedings of the 25th International Symposium on Software
  Testing and Analysis}, ISSTA 2016, 2016.

\bibitem{Baudry2006}
B.~Baudry, F.~Fleurey, and Y.~Le~Traon.
\newblock {Improving Test Suites for Efficient Fault Localization}.
\newblock In {\em {28th International Conference on Software Engineering (ICSE
  06)}}, Shanghai, China, 2006. {ACM}.
\newblock selection : 9\%.

\bibitem{beizer:testing}
B.~Beizer.
\newblock {\em Software Testing Techniques}.
\newblock Van {N}ostrand {R}einhold, 1990.

\bibitem{setTimeLimit}
H.~Bengtsson.
\newblock Linux settimelimit bug description.

\bibitem{bertolino:testing-fose07}
A.~Bertolino.
\newblock Software testing research: {A}chievements, challenges, dreams.
\newblock In L.~Briand and A.~Wolf, editors, {\em Future of Software
  Engineering 2007}, Los Alamitos, California, USA, 2007. {IEEE} {C}omputer
  {S}ociety {P}ress.

\bibitem{binder:design}
R.~V. Binder.
\newblock Design for testability in object--oriented systems.
\newblock {\em Communications of the ACM}, 37(9):87--101, 1994.

\bibitem{chandra:how}
S.~Chandra and P.~M. Chen.
\newblock How fail-stop are faulty programs?
\newblock In {\em 28th IEEE Symposium on Fault Tolerant Computing Systems
  ({FTCS-28})}, pages 240--249, June 1998.

\bibitem{DaranT96}
M.~Daran and P.~Th{\'e}venod-Fosse.
\newblock Software error analysis: A real case study involving real faults and
  mutations.
\newblock In {\em Proceedings of the 1996 ACM SIGSOFT International Symposium
  on Software Testing and Analysis}, ISSTA '96, pages 158--171, New York, NY,
  USA, 1996. ACM.

\bibitem{fraser:evosuite}
G.~Fraser and A.~Arcuri.
\newblock {EvoSuite}: automatic test suite generation for object-oriented
  software.
\newblock In {\em $8^{th}$ European Software Engineering Conference and the
  {ACM SIGSOFT} Symposium on the Foundations of Software Engineering ({ESEC/FSE
  '11})}, pages 416--419. ACM, September 5th - 9th 2011.

\bibitem{grottke:classification}
M.~Grottke and K.~Trivedi.
\newblock A classification of software faults.
\newblock In {\em Supplemental Proc. Sixteenth International {IEEE} Symposium
  on Software Reliability Engineering}, pages 4.19 -- 4.20, 2005.

\bibitem{Gunawi2008}
H.~S. Gunawi, C.~Rubio-Gonz\'{a}lez, A.~C. Arpaci-Dusseau, R.~H. Arpaci-Dussea,
  and B.~Liblit.
\newblock Eio: Error handling is occasionally correct.
\newblock In {\em Proceedings of the 6th USENIX Conference on File and Storage
  Technologies}, FAST'08, pages 14:1--14:16, Berkeley, CA, USA, 2008. USENIX
  Association.

\bibitem{gunawi:eio}
H.~S. Gunawi, C.~Rubio-Gonz{\'a}lez, A.~C. Arpaci-Dusseau, R.~H.
  Arpaci-Dusseau, and B.~Liblit.
\newblock {EIO}: Error handling is occasionally correct.
\newblock In M.~Baker and E.~Riedel, editors, {\em 6th {USENIX} Conference on
  File and Storage Technologies ({FAST} 2008)}, pages 207--222, 2008.

\bibitem{mhetal:tse-flag}
M.~Harman, L.~Hu, R.~M. Hierons, J.~Wegener, H.~Sthamer, A.~Baresel, and
  M.~Roper.
\newblock Testability transformation.
\newblock {\em IEEE Transactions on Software Engineering}, 30(1):3--16, Jan.
  2004.

\bibitem{yjmh:analysis}
Y.~Jia and M.~Harman.
\newblock An analysis and survey of the development of mutation testing.
\newblock {\em {IEEE} Transactions on Software Engineering}, 37(5):649 -- 678,
  September--October 2011.

\bibitem{Jones2004}
J.~A. Jones.
\newblock Fault localization using visualization of test information.
\newblock In {\em Proceedings. 26th International Conference on Software
  Engineering}, pages 54--56, May 2004.

\bibitem{just:major}
R.~Just.
\newblock The {M}ajor mutation framework: Efficient and scalable mutation
  analysis for {J}ava.
\newblock In {\em Proceedings of the International Symposium on Software
  Testing and Analysis (ISSTA)}, pages 433--436, San Jose, CA, USA, July~23--25
  2014.

\bibitem{just:defects4j}
R.~Just, D.~Jalali, and M.~D. Ernst.
\newblock Defects4j: A database of existing faults to enable controlled testing
  studies for java programs.
\newblock In {\em Proceedings of the 2014 International Symposium on Software
  Testing and Analysis}, ISSTA 2014, pages 437--440, New York, NY, USA, 2014.
  ACM.

\bibitem{Just:2014:MVS}
R.~Just, D.~Jalali, L.~Inozemtseva, M.~D. Ernst, R.~Holmes, and G.~Fraser.
\newblock Are mutants a valid substitute for real faults in software testing?
\newblock In {\em Proceedings of the 22Nd ACM SIGSOFT International Symposium
  on Foundations of Software Engineering}, FSE 2014, 2014.

\bibitem{just:fse}
R.~Just, D.~Jalali, L.~Inozemtseva, M.~D. Ernst, R.~Holmes, and G.~Fraser.
\newblock Are mutants a valid substitute for real faults in software testing?
\newblock In {\em International Symposium on Foundations of Software
  Engineering ({FSE})}, pages 654--665, 2014.

\bibitem{JustJIEHF14}
R.~Just, D.~Jalali, L.~Inozemtseva, M.~D. Ernst, R.~Holmes, and G.~Fraser.
\newblock Are mutants a valid substitute for real faults in software testing?
\newblock In {\em Proceedings of the 22nd {ACM} {SIGSOFT} International
  Symposium on Foundations of Software Engineering ({FSE})}, pages 654--665,
  2014.

\bibitem{li2012using}
Y.~Li and C.~Liu.
\newblock Using cluster analysis to identify coincidental correctness in fault
  localization.
\newblock In {\em Computational and Information Sciences (ICCIS), 2012 Fourth
  International Conference on}, pages 357--360. IEEE, 2012.

\bibitem{Lu:2016:RTP}
Y.~Lu, Y.~Lou, S.~Cheng, L.~Zhang, D.~Hao, Y.~Zhou, and L.~Zhang.
\newblock How does regression test prioritization perform in real-world
  software evolution?
\newblock In {\em Proceedings of the 38th International Conference on Software
  Engineering}, ICSE '16, 2016.

\bibitem{Ma:2015:GRT}
L.~Ma, C.~Artho, C.~Zhang, H.~Sato, J.~Gmeiner, and R.~Ramler.
\newblock Grt: Program-analysis-guided random testing (t).
\newblock In {\em 2015 30th IEEE/ACM International Conference on Automated
  Software Engineering (ASE)}, 2015.

\bibitem{Mart:2016:APR}
M.~Martinez and M.~Monperrus.
\newblock Astor: A program repair library for java (demo).
\newblock In {\em Proceedings of the 25th International Symposium on Software
  Testing and Analysis}, ISSTA 2016, 2016.

\bibitem{masri2009empirical}
W.~Masri, R.~Abou-Assi, M.~El-Ghali, and N.~Al-Fatairi.
\newblock An empirical study of the factors that reduce the effectiveness of
  coverage-based fault localization.
\newblock In {\em Proceedings of the 2nd International Workshop on Defects in
  Large Software Systems: Held in conjunction with the ACM SIGSOFT
  International Symposium on Software Testing and Analysis (ISSTA 2009)}, pages
  1--5. ACM, 2009.

\bibitem{masri2010}
W.~Masri and R.~A. Assi.
\newblock Cleansing test suites from coincidental correctness to enhance
  fault-localization.
\newblock In {\em Software Testing, Verification and Validation (ICST), 2010
  Third International Conference on}, pages 165--174. IEEE, 2010.

\bibitem{MasriA14}
W.~Masri and R.~A. Assi.
\newblock Prevalence of coincidental correctness and mitigation of its impact
  on fault localization.
\newblock {\em {ACM} Trans. Softw. Eng. Methodol.}, 23(1):8:1--8:28, 2014.

\bibitem{mcminn:co-tetra}
P.~McMinn.
\newblock Search-based failure discovery using testability transformations to
  generate pseudo-oracles.
\newblock In F.~Rothlauf, editor, {\em Genetic and Evolutionary Computation
  Conference ({GECCO} 2009)}, pages 1689--1696, Montreal, Qu{\'e}bec, Canada,
  2009. {ACM}.

\bibitem{miao2012}
Y.~Miao, Z.~Chen, S.~Li, Z.~Zhao, and Y.~Zhou.
\newblock Identifying coincidental correctness for fault localization by
  clustering test cases.
\newblock In {\em SEKE}, pages 267--272, 2012.

\bibitem{Pea:2017:EIF}
S.~Pearson, J.~Campos, R.~Just, G.~Fraser, R.~Abreu, M.~D. Ernst, D.~Pang, and
  B.~Keller.
\newblock Evaluating and improving fault localization.
\newblock In {\em Proceedings of the 39th International Conference on Software
  Engineering}, ICSE '17, 2017.

\bibitem{randell:fandf}
B.~Randell.
\newblock On failures and faults.
\newblock In K.~Araki, S.~Gnesi, and D.~Mandrioli, editors, {\em International
  Symposium of Formal Methods Europe ({FME})}, volume 2805 of {\em Lecture
  Notes in Computer Science}, pages 18--39. Springer, 2003.

\bibitem{Rubio-Gonzalez2009PLDI}
C.~Rubio-Gonz\'{a}lez, H.~S. Gunawi, B.~Liblit, R.~H. Arpaci-Dusseau, and A.~C.
  Arpaci-Dusseau.
\newblock Error propagation analysis for file systems.
\newblock In {\em Proceedings of the 30th ACM SIGPLAN Conference on Programming
  Language Design and Implementation}, PLDI '09, pages 270--280, New York, NY,
  USA, 2009. ACM.

\bibitem{Rubio-Gonzalez2009SigPlan}
C.~Rubio-Gonz\'{a}lez, H.~S. Gunawi, B.~Liblit, R.~H. Arpaci-Dusseau, and A.~C.
  Arpaci-Dusseau.
\newblock Error propagation analysis for file systems.
\newblock {\em SIGPLAN Not.}, 44(6):270--280, June 2009.

\bibitem{Sham:2015:AUT}
S.~Shamshiri.
\newblock Automated unit test generation for evolving software.
\newblock In {\em Proceedings of the 2015 10th Joint Meeting on Foundations of
  Software Engineering}, ESEC/FSE 2015, 2015.

\bibitem{ShashaZhang}
D.~Shasha and K.~Zhang.
\newblock {\em Approximate tree pattern matching}, pages 341--371.
\newblock Oxford University Press, 1997.

\bibitem{StaatsGH12}
M.~Staats, G.~Gay, and M.~P.~E. Heimdahl.
\newblock Automated oracle creation support, or: How {I} learned to stop
  worrying about fault propagation and love mutation testing.
\newblock In {\em 34th International Conference on Software Engineering, {ICSE}
  2012, June 2-9, 2012, Zurich, Switzerland}, pages 870--880, 2012.

\bibitem{Voas1991}
J.~Voas, L.~Morell, and K.~Miller.
\newblock Predicting where faults can hide from testing.
\newblock {\em IEEE Software}, 8(2):41--48, March 1991.

\bibitem{Voas1992}
J.~M. Voas.
\newblock Pie: A dynamic failure-based technique.
\newblock {\em IEEE Trans. Softw. Eng.}, 18(8):717--727, Aug. 1992.

\bibitem{VoasMiller1994}
J.~M. Voas and K.~W. Miller.
\newblock Putting assertions in their place.
\newblock In {\em Proceedings of 1994 IEEE International Symposium on Software
  Reliability Engineering}, pages 152--157, Nov 1994.

\bibitem{voas:testability}
J.~M. Voas and K.~W. Miller.
\newblock Software testability: The new verification.
\newblock {\em {IEEE} Software}, 12(3):17--28, May 1995.

\bibitem{WangCCZ09}
X.~Wang, S.~Cheung, W.~K. Chan, and Z.~Zhang.
\newblock Taming coincidental correctness: Coverage refinement with context
  patterns to improve fault localization.
\newblock In {\em 31st International Conference on Software Engineering, {ICSE}
  2009, May 16-24, 2009, Vancouver, Canada, Proceedings}, pages 45--55, 2009.

\bibitem{wong:adaptive}
E.~Wong and B.~Cukic.
\newblock {\em Adaptive Control Approach for Software Quality Improvement}.
\newblock World scientific Series on Software Engineering and Knowledge
  Engineering, 2011.
\newblock Volume 20.

\bibitem{Wong2007}
W.~E. Wong, Y.~Qi, L.~Zhao, and K.~Y. Cai.
\newblock Effective fault localization using code coverage.
\newblock In {\em 31st Annual International Computer Software and Applications
  Conference (COMPSAC 2007)}, volume~1, pages 449--456, July 2007.

\bibitem{Xin:2017:ITP}
Q.~Xin and S.~P. Reiss.
\newblock Identifying test-suite-overfitted patches through test case
  generation.
\newblock In {\em Proceedings of the 26th ACM SIGSOFT International Symposium
  on Software Testing and Analysis}, ISSTA 2017, 2017.

\bibitem{XiongH0ZZL15}
Y.~Xiong, D.~Hao, L.~Zhang, T.~Zhu, M.~Zhu, and T.~Lan.
\newblock Inner oracles: input-specific assertions on internal states.
\newblock In {\em Proceedings of the 2015 10th Joint Meeting on Foundations of
  Software Engineering, {ESEC/FSE}}, pages 902--905, 2015.

\bibitem{Xiong:2017:PCS}
Y.~Xiong, J.~Wang, R.~Yan, J.~Zhang, S.~Han, G.~Huang, and L.~Zhang.
\newblock Precise condition synthesis for program repair.
\newblock In {\em Proceedings of the 39th International Conference on Software
  Engineering}, ICSE '17, 2017.

\bibitem{XuePN14}
X.~Xue, Y.~Pang, and A.~S. Namin.
\newblock Trimming test suites with coincidentally correct test cases for
  enhancing fault localizations.
\newblock In {\em {IEEE} 38th Annual Computer Software and Applications
  Conference, {COMPSAC} 2014, Vasteras, Sweden, July 21-25, 2014}, pages
  239--244, 2014.

\bibitem{zheng2013}
Z.~Zheng, Y.~Gao, P.~Hao, and Z.~Zhang.
\newblock Coincidental correctness: An interference or interface to successful
  fault localization?
\newblock In {\em Software Reliability Engineering Workshops (ISSREW), 2013
  IEEE International Symposium on}, pages 114--119. IEEE, 2013.

\end{thebibliography}

\end{document}